\documentclass[%
reprint,
superscriptaddress,
showpacs,
preprintnumbers,
 amsmath,amssymb,
pra,
longbibliography
]{revtex4-1}

\usepackage{graphicx}% Include figure files
\usepackage{dcolumn}% Align table columns on decimal point
\usepackage{bm}% bold math
\usepackage{setspace}	% required for `\spacing' (yatex added)
\usepackage{color}
\usepackage{amsmath}	% required for `\align' (yatex added)
\usepackage{upgreek}
\DeclareGraphicsExtensions{.eps}
\def\ii{{\mathrm{i}}}

\def\ee{{\mathrm{e}}}
\def\dd{{\mathrm{d}}}

\def\no{{\nonumber}} %\nonumber

\def\bra#1{\langle #1|}

\def\ket#1{|#1\rangle}

\def\bracketi#1#2{\langle #1 | #2 \rangle}
\def\bracketii#1#2#3{\langle #1 | #2| #3\rangle}

\def\sub#1{_\mathrm{#1}} %subscript
\def\HH{\mathrm{H}}
\def\VV{\mathrm{V}}
\def\DD{\mathrm{D}}
\def\AA{\mathrm{A}}
\def\RR{\mathrm{R}}
\def\LL{\mathrm{L}}

\begin{document}

%\preprint{Ver.5.1}

\title{
Direct measurement of ultrafast temporal wavefunctions 
}

% Force line breaks with \\

\author{Kazuhisa Ogawa}
\email{ogawak@ist.hokudai.ac.jp}
\affiliation{% 
Graduate School of Information Science and Technology, Hokkaido University, Sapporo 060-0814, Japan
}%
 
\author{Takumi Okazaki}
\affiliation{% 
Graduate School of Information Science and Technology, Hokkaido University, Sapporo 060-0814, Japan
}%

\author{Hirokazu Kobayashi}
\affiliation{% 
School of System Engineering, Kochi University of Technology, Kochi 782-8502, Japan
}%

\author{Toshihiro Nakanishi}
\affiliation{% 
Department of Electronic Science and Engineering, Kyoto University, Kyoto 615-8510, Japan
}%

\author{Akihisa Tomita}
\affiliation{% 
Graduate School of Information Science and Technology, Hokkaido University, Sapporo 060-0814, Japan
}%

\date{\today}% It is always \today, today,

\begin{abstract}
The large capacity and robustness of information encoding in the temporal mode of photons is important in quantum information processing, in which characterizing temporal quantum states with high usability and time resolution is essential.
We propose and demonstrate a direct measurement method of temporal complex wavefunctions for weak light at a single-photon level with subpicosecond time resolution.
% The outcome of the local measurement for one part of the wavefunction directly corresponds to its complex value at that part.
Our direct measurement is realized by ultrafast metrology of the interference between the light under test and self-generated monochromatic reference light; no external reference light or complicated post-processing algorithms are required.
Hence, this method is versatile and potentially widely applicable for temporal state characterization. 
\end{abstract}

% \pacs{42.30.Kq, 42.50.St, 42.50.Xa}
\maketitle

%\tableofcontents

\begin{spacing}{1.2}

\section{Introduction}

The temporal--spectral mode of photons offers an attractive platform for quantum information processing in terms of a large capacity due to its high dimensionality and robustness in fiber and waveguide transmission.
To date, many applications using the temporal--spectral mode have been proposed and realized in quantum information processing fields such as quantum computation, quantum cryptography, and quantum metrology \cite{PhysRevLett.113.130502,nunn2013large,PhysRevA.87.062322,PhysRevLett.112.133602,roslund2014wavelength,PhysRevX.5.041017,PhysRevLett.101.123601,jian2012real,PhysRevLett.111.150501,ryczkowski2016ghost}.
In these applications, the full characterization of quantum states, i.e.,~complex wavefunctions, is crucial for developing reliable quantum operations.
In addition, temporal-mode characterization for high-speed and precise processing often requires ultrafast time resolution, such as on the subpicosecond scale.

\begin{figure}
\centering
\includegraphics[width=8.5cm]{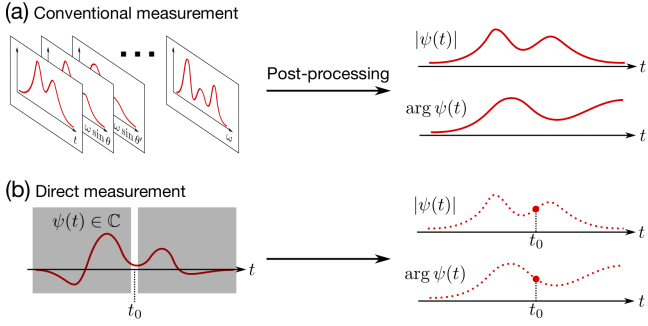}
\caption{
Comparison of the concepts of the conventional and direct measurement methods.
(a) In conventional measurement methods, measurements of the entire wavefunction are usually performed first, and then the measurement data are post-processed to reconstruct the entire wavefunction.
(b) In the direct measurement method, the measurement data at the time $t_0$ directly correspond to the complex value $\psi(t_0)$ of the wavefunction $\psi(t)$ at $t_0$.
}
\label{fig:0}
\end{figure}

Several established methods, such as frequency-resolved optical gating (FROG) and spectral phase interferometry for direct electric field reconstruction (SPIDER), are well known for measuring the temporal--spectral mode of classical light \cite{walmsley2009characterization}. 
These methods, however, utilize the nonlinear optical processes of the light under test, which are difficult to observe for weak light at the single-photon level.
In recent years, various methods for characterizing the temporal--spectral mode of quantum light have been demonstrated, such as single photons and entangled photon pairs \cite{PhysRevLett.120.213601,PhysRevLett.109.053602,qin2015complete,PhysRevApplied.10.054011,PhysRevLett.99.123601,PhysRevA.100.042317,PhysRevLett.114.010401,PhysRevLett.121.083602,davis2020measuring,PhysRevA.98.023840,PhysRevA.100.033834,thiel2020single}, and some have achieved ultrafast (subpicosecond) time resolution \cite{PhysRevLett.120.213601,PhysRevLett.109.053602,PhysRevLett.99.123601,PhysRevLett.121.083602,davis2020measuring,PhysRevA.98.023840,PhysRevA.100.033834,thiel2020single}.
While these methods differ in the details of their measurement procedures, they have a common procedure to reconstruct the form of the wavefunction: projective measurements for the entire temporal (or spectral or other basis) wavefunction have to be performed first, and then the measurement data is post-processed, as shown in Fig.~\ref{fig:0}(a).
In other words, even for acquiring only one part of the wavefunction, measurement of the entire wavefunction is essential.
Each set of measurement data before post-processing contains partial information of the wavefunction but is not itself the wavefunction.

As a more suitable measurement method for the form of the wavefunction, \textit{direct measurement} \cite{lundeen2011direct} is the focus of this study.
The direct measurement of a wavefunction $\psi(t)$ is defined as the measurement that can reconstruct the complex value $\psi(t_0)$ only using the measurement data at the point $t=t_0$, as shown in Fig.~\ref{fig:0}(b); that is, the measurement data at $t_0$ \textit{directly} correspond to the complex value $\psi(t_0)$.
Direct measurement was first demonstrated for the transverse spatial wavefunction of single photons \cite{lundeen2011direct} using a technique called weak measurement \cite{PhysRevLett.60.1351}, and then for wavefunctions and density matrices in various degrees of freedom \cite{PhysRevLett.108.070402,malik2014direct,salvail2013full,shi2015scan,PhysRevLett.117.120401}. 
While direct measurement was introduced to give the operational meaning of the complex-valued wavefunction, it also provides a practical advantage of requiring only one measurement basis.
Although direct measurement using weak measurement has drawbacks in its approximation error and low efficiency due to the nature of weak measurement, in recent years it has been reported that direct measurement can also be realized using strong (projection) measurement both theoretically \cite{PhysRevA.91.052109,PhysRevLett.116.040502,ogawa2019framework} and experimentally \cite{PhysRevLett.118.010402,PhysRevLett.121.230501}.
Therefore, applying direct measurement using strong measurement to the temporal wavefunction of photons can provide a practical characterization method for temporal wavefunctions, which avoids the requirement of post-processing the measurement data of the entire wavefunction.

In this paper, we propose a direct measurement method of temporal complex wavefunctions that can be performed for weak light at a single-photon level with subpicosecond time resolution. 
Our direct measurement is realized by ultrafast metrology (time gate measurement) of the interference between the light under test and the self-generated monochromatic reference light with several phase differences.
This mechanism is simple compared to other measurement methods of the temporal--spectral mode of quantum light; that is, it does not require external reference light or complicated post-processing of the measurement data.
We also experimentally demonstrate this direct measurement method of the temporal wavefunction of light at a single-photon level and examine the validity of the measurement results.

\section{Theory}

The proposed method for direct measurement of the temporal wavefunction is based on our previous study \cite{ogawa2019framework}.
The wavefunction under test $\psi(t)$ is the temporal representation of the pulse-mode state $\ket{\psi}$, and its spectral representation $\tilde{\psi}(\omega)$ is given by the Fourier transform of $\psi(t)$.
$\psi(t)$ can be represented by the product of the complex-valued envelope function $\psi\sub{env}(t)$ and the carrier term $\ee^{-\ii\omega_0 t}$ as $\psi(t)=\psi\sub{env}(t)\ee^{-\ii\omega_0 t}$, where $\omega_0$ is the reference carrier frequency.
%define  as $\psi\sub{env}(t):=\psi(t)\ee^{\ii\omega_0 t}$, 
We assume that $\omega_0$ is known and then consider measuring $\psi\sub{env}(t)$ instead of $\psi(t)$.
The Fourier transform of $\psi\sub{env}(t)$, $\tilde{\psi}\sub{env}(\omega)$, satisfies the relation $\tilde{\psi}\sub{env}(\omega)=\tilde{\psi}(\omega+\omega_0)$.

The basic mechanism common to most direct measurements \cite{lundeen2011direct,PhysRevLett.108.070402,malik2014direct,salvail2013full,shi2015scan,PhysRevLett.117.120401,PhysRevA.91.052109,PhysRevLett.116.040502,ogawa2019framework,PhysRevLett.118.010402,PhysRevLett.121.230501,ogawa2019framework} is the interference between the signal wavefunction under test $\psi(t)=\psi\sub{env}(t)\ee^{-\ii\omega_0 t}$ and a self-generated uniform reference wave $\psi_0\ee^{-\ii\omega_0t}$ with four phase differences $0$, $\pi/2$, $\pi$, and $3\pi/2$.
The probability that their superposition state is projected onto time $t$ and phase difference $\theta$ is given by $P(t,\theta)=|\psi(t)+\ee^{-\ii(\omega_0t+\theta)}\psi_0|^2=|\psi\sub{env}(t)+\ee^{-\ii\theta}\psi_0|^2$.
% \begin{align}
% P_0(t)=|\psi\sub{env}(t)+c|^2,&\quad
% P_{\pi/2}(t)=|\psi\sub{env}(t)+\ii c|^2,\\
% P_{\pi}(t)=|\psi\sub{env}(t)-c|^2,&\quad
% P_{3\pi/2}(t)=|\psi\sub{env}(t)-\ii c|^2.
% \end{align}
The differences between $P(t,0)$ and $P(t,\pi)$ and between $P(t,\pi/2)$ and $P(t,3\pi/2)$ give the real and imaginary parts of $\psi\sub{env}(t)$, respectively:
\begin{align}
P(t,0)-P(t,\pi)&\propto\mathrm{Re}[\psi\sub{env}(t)],\\
P(t,\pi/2)-P(t,3\pi/2)&\propto\mathrm{Im}[\psi\sub{env}(t)].
\end{align}
Their proportional coefficients are equal and can be determined by the normalization condition of the wavefunction.

% 我々が行う波動関数の直接測定法は, 文献\cite{ogawa2019framework}において提案されたものである. 
% 本実験ではその測定法を時間波動関数に適用した. 
% 被状態$\ket{\psi}$の時間波動関数は$\bracketi{t}{\psi}=\psi(t)=\psi\sub{env}(t)\ee^{-\ii\omega_0t}$と表され, ここで$\omega_0$はキャリア中心周波数, $\psi\sub{env}(t)$は複素包絡線関数を表す. 
% 本実験では$\omega_0$は既知の固定値とし, $\psi(t)$の代わりに$\psi\sub{env}(t)$の直接測定を考える.  
% 本手法では, 被測定系に加えてqubitプローブ系を用いる. 
% qubitプローブ系として, 我々は水平偏光状態$\ket{\HH}$, 垂直偏光状態$\ket{\VV}$で張られる偏光状態自由度を用いる. 
% $\ket{\HH}$と$\ket{\VV}$の等しい重ね合わせ状態4つを次のように定義する: diagonal $\ket{\DD}:=(\ket{\HH}+\ket{\VV})/\sqrt{2}$, anti-diagonal $\ket{\AA}:=(\ket{\HH}-\ket{\VV})/\sqrt{2}$, right-circular $\ket{\RR}:=(\ket{\HH}+\ii\ket{\VV})/\sqrt{2}$, left-circular $\ket{\LL}:=(\ket{\HH}-\ii\ket{\VV})/\sqrt{2}$. 

\begin{figure}
\centering
\includegraphics[width=8.5cm]{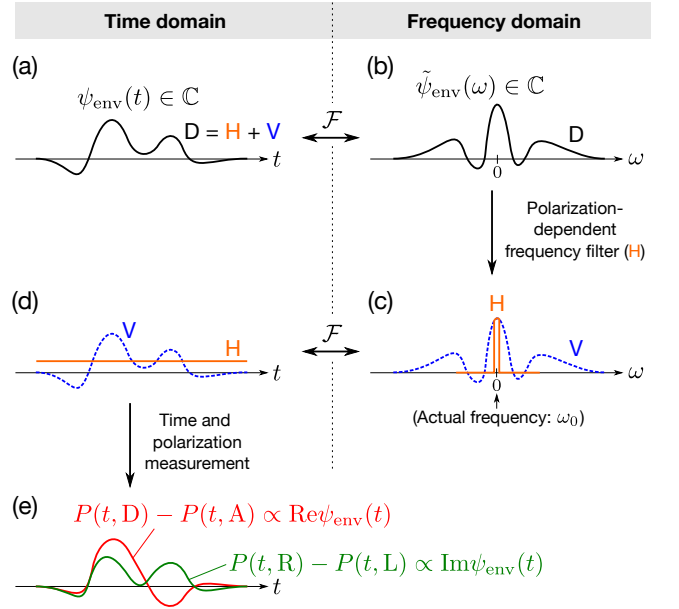}
\caption{Procedure of direct measurement of the temporal wavefunction.
(a), (b) Temporal and spectral representations of the pulse-mode state $\ket{\psi}$, respectively.
Its polarization mode is set to the diagonal state $\ket{\DD}$.
% (b) Spectral representation of $\psi\sub{env}(t)$, $\tilde{\psi}\sub{env}(\omega)$, which is the Fourier transform of $\psi\sub{env}(t)$.
%  $\omega'$ is the difference from the carrier central frequency $\omega_0$.
(c) Wavefunction after applying the polarization-dependent frequency filter at $\omega=0$ (the actual frequency is $\omega_0$) to the wavefunction of (b).
Only the frequency $\omega=0$ component remains for the horizontally polarized light.
(d) Temporal representation of the wavefunction of (c).
The horizontally polarized component has an almost uniform distribution, which serves as a reference for the magnitude and phase of the vertically polarized temporal wavefunction $\psi\sub{env}(t)$.
(e) Real and imaginary parts of $\psi\sub{env}(t)$, which are reconstructed by combining the projection probability distributions $P(t,\DD)$, $P(t,\AA)$, $P(t,\RR)$, and $P(t,\LL)$ of time and polarization measurement for the wavefunction of (d).
% 我々が提案する時間波動関数の直接測定の手順. 
% (a)測定対象である時間波動関数$\psi\sub{env}(t)$. 偏光状態を斜め45$\degree$偏光状態 (D)に準備する. 
% (b) (a)で表される波動関数の周波数領域における表示. 
% (c) (b)で表される波動関数に偏光依存周波数フィルタをさせた後の状態. 
% 水平偏光(H)成分において周波数$\omega_0$の部分だけ取り出される. 
% (d) (c)で表される波動関数の時間領域における表示. 
% 水平偏光(H)成分はほぼ一様な分布であり, これが垂直偏光(V)成分の波動関数$\psi\sub{env}(t)$に対する大きさと位相の基準の役割を果たす. 
% (e) 時間, 偏光についての射影確率分布$P(t,\DD)$, $P(t,\AA)$, $P(t,\RR)$, $P(t,\LL)$を組み合わせると, 被測定波動関数$\psi\sub{env}(t)$の実部と虚部が再構成できる. 
}
\label{fig:1}
\end{figure}

To realize the above mechanism, our direct measurement method \cite{ogawa2019framework} uses a qubit (two-state quantum system) probe mode
% in addition to the temporal--spectral mode under test 
to prepare the four phase differences; we utilize the polarization mode of the photons spanned by the horizontal and vertical states $\ket{\HH}$ and $\ket{\VV}$. 
We define the four polarization states as follows: diagonal $\ket{\DD}:=(\ket{\HH}+\ket{\VV})/\sqrt{2}$, anti-diagonal $\ket{\AA}:=(\ket{\HH}-\ket{\VV})/\sqrt{2}$, right-circular $\ket{\RR}:=(\ket{\HH}+\ii\ket{\VV})/\sqrt{2}$, and left-circular $\ket{\LL}:=(\ket{\HH}-\ii\ket{\VV})/\sqrt{2}$. 
The procedure of our direct measurement of the temporal wavefunction is shown in Fig.~\ref{fig:1}. 
Let the initial state be $\ket{\varPsi_0}:=\ket{\psi}\ket{\DD}=\ket{\psi}(\ket{\HH}+\ket{\VV})/\sqrt{2}$.
The temporal and spectral representations of $\ket{\varPsi_0}$ are shown in Figs.~\ref{fig:1}(a) and (b), respectively. 
First, we extract the frequency $\omega=0$ component (the actual frequency is $\omega_0$) from the horizontally polarized light using a polarization-dependent frequency filter.
This operation is ideally described by the projection operator $\ket{\omega_0}\bra{\omega_0}\otimes\ket{\HH}\bra{\HH}+\hat{1}\otimes\ket{\VV}\bra{\VV}$, and the unnormalized state after the projection is given by $\ket{\varPsi_1}:=(\ket{\omega_0}\bracketi{\omega_0}{\psi}\ket{\HH}+\ket{\psi}\ket{\VV})/\sqrt{2}$.
Second, we perform projection measurements of time and polarization for $\ket{\varPsi_1}$.
The projections onto D, A, R, L polarizations correspond to the preparations of the four phase differences $0$, $\pi$, $\pi/2$, and $3\pi/2$, respectively.
The projection operator onto time $t$ and polarization $\phi$ is described as $\ket{t}\bra{t}\otimes\ket{\phi}\bra{\phi}$, and its projection probability is given by $P(t,\phi)=\bracketii{\varPsi_1}{(\ket{t}\bra{t}\otimes\ket{\phi}\bra{\phi})}{\varPsi_1}/\bracketi{\varPsi_1}{\varPsi_1}$.
Using $P(t,\phi)$ for $\phi=\DD$, $\AA$, $\RR$, and $\LL$, the real and imaginary parts of $\psi\sub{env}(t)$ are obtained as
\begin{align}
P(t,\DD)-P(t,\AA)&\propto
\mathrm{Re}[\bracketi{\psi}{\omega_0}\bracketi{\omega_0}{t}\bracketi{t}{\psi}]\no\\
&\propto
\mathrm{Re}[\ee^{\ii\omega_0t}\psi(t)]
=\mathrm{Re}[\psi\sub{env}(t)],\label{eq:1}\\
P(t,\RR)-P(t,\LL)&\propto
\mathrm{Im}[\psi\sub{env}(t)],\label{eq:2}
\end{align}
where $\bracketi{\psi}{\omega_0}$ is a constant that does not depend on $t$ and $\bracketi{\omega_0}{t}=\ee^{\ii\omega_0t}/\sqrt{2\pi}$.

Here, we emphasize the following two points.
First, our measurement method satisfies the definition of direct measurement mentioned previously.
% , although it seems different from the direct measurement using weak measurement \cite{lundeen2011direct}.
Indeed, to obtain the complex value of $\psi\sub{env}(t_0)$, this measurement method requires only the four projection probabilities $P(t_0,\phi)$ ($\phi=\DD,\AA,\RR,\LL$) at time $t_0$.
Second, our direct measurement method is more accurate and efficient than conventional direct measurement methods using weak measurement \cite{lundeen2011direct,PhysRevLett.108.070402,malik2014direct,salvail2013full,shi2015scan,PhysRevLett.117.120401}.
% as we also mentioned before, the essential point for the direct measurement of the wavefunction is the interference between the signal and the self-generated uniform reference wave.
Our measurement method causes interference between the signal and the self-generated uniform reference wave using the polarization-dependent frequency filter (projection measurement) instead of weak measurement.
Therefore, our method can avoid the approximation error and low measurement efficiency associated with weak measurement.
We note that the polarization degree of freedom, which is used to provide the four phase differences in the interference, can be replaced by another degree of freedom such as path mode when the polarization mode is already used or unstable for use.

\section{Experiments}\label{sec:experiments}

\begin{figure*}
\includegraphics[width=18cm]{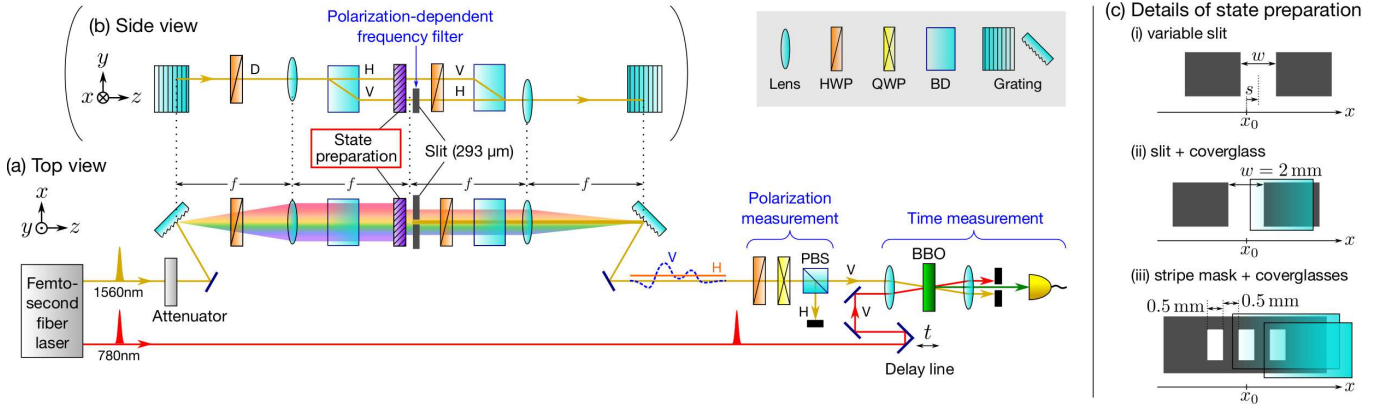}
\caption{Experimental setup.
(a) Top view of the whole system. 
(b) Side view of the 4-$f$ system.
HWP: half-wave plate, QWP: quarter-wave plate, BD: beam displacer, PBS: polarizing beam splitter, BBO: $\upbeta$-BaB$_2$O$_4$ crystal.
(c) Details of state preparation.
(i) Variable slit ($w$: gap width, $s$: displacement of the gap center from $x_0$, $x_0$: center position of the slit for polarization-dependent frequency filtering).
(ii) Slit ($w=2\,\mathrm{mm}$, $s=0\,\mathrm{mm}$) and coverglass ($170\pm 5\,\mathrm{mm}$ thickness).
(iii) Stripe mask (0.5\,mm gap) and two coverglasses.
}\label{fig:2}
\end{figure*}

We demonstrate the direct measurement of the temporal wavefunction using the measurement system shown in Fig.~\ref{fig:2}.
The femtosecond fiber laser (Menlo Systems C-Fiber 780) emits two pulsed light beams with central wavelengths 1560\,nm and 780\,nm in synchronization (repetition rate 100\,MHz).
% The fs fiber laser emits synchronized two pulsed beams with central wavelengths 1560\,nm and 780\,nm.
The 1560\,nm beam is used as the signal under test, and the 780\,nm beam [76.5\,mW, 79.2\,fs full width at half maximum (FWHM)] as the gate pulse for the time gate measurement \footnote{While the 780\,nm beam is not only synchronized but also has coherence with the 1560\,nm beam, this coherence is not necessary for the time gate measurement.}.
We prepare the signal power in the following two conditions using the attenuator: the classical-light (CL) condition, in which the average photon number is 366\,photons/pulse ($4.69\,\mathrm{nW}$); and the single-photon-level (SPL) condition, in which the average photon number is 0.58\,photons/pulse ($7.47\,\mathrm{pW}$) and the probability of one or fewer photons per pulse is 0.885.
The SPL condition is used to demonstrate that our direct measurement system works even for a signal as weak as a single-photon level.

The 1560\,nm beam then enters the 4-$f$ system composed of the gratings (600\,grooves/mm) and lenses (focal length $f=300\,\mathrm{mm}$).
At the center of the 4-$f$ system, the spectral distribution is mapped onto the transverse spatial distribution, where state preparation followed by polarization-dependent frequency filtering is performed.
As seen in Fig.~\ref{fig:2}(b), two beam displacers (BDs) are set in the 4-$f$ system to divide the optical path according to the polarization; the polarization-dependent frequency filter is realized by inserting a slit (293\,$\upmu$m width) in one of the paths.
In contrast, the state preparation before the slit is performed equally for the two beams.
After the state preparation followed by polarization-dependent frequency filtering, the polarizations of the two beams are exchanged by the half-wave plate (HWP) and then combined by the second BD so that the two optical path lengths are equal.

In the state preparation, we prepare the three types of states shown in Fig.~\ref{fig:2}(c).
A variable slit with gap width $w$ and displacement $s$ is used to quantitatively evaluate the measured temporal wavefunction.
The coverglass is used to cause a phase change.
As the magnitude of the phase change depends sensitively on the inclination of the coverglass, we assume that this magnitude is unknown.
The combination of the stripe mask and coverglasses is used to demonstrate the direct measurement of a complicated wavefunction.

After the 4-$f$ system, the beam is projected onto one of the D, A, R, or L polarizations by the HWP, quarter-wave plate, and polarizing beam splitter.
Subsequently, the beam is projected onto time $t$ by the time gate measurement, which is realized by sum-frequency generation (SFG) of the signal beam and the 780\,nm gate pulse with delay $t$.
In SFG, these two beams are focused on the $\upbeta$-BaB$_2$O$_4$ crystal by the lens ($f=50\,\mathrm{mm}$), and their sum-frequency light (520\,nm wavelength) is emitted at an intensity proportional to the product of the two input temporal intensities.
By scanning the delay of the gate pulse $t$, sum-frequency light with an intensity proportional to the time intensity distribution of the signal light is extracted.
% 十分に時間幅の狭いゲートパルスのディレイ$t$をスキャンすれば、
% シグナル光の瞬時強度に比例した強度の和周波光が取り出せる
Finally, the sum-frequency light is spatially and spectrally filtered to remove the stray light (not shown in the figure) and then detected by a single-photon counting module (Laser Components COUNT-NIR).

For comparison, we additionally perform intensity (projection) measurements in time and frequency for the state under test in the CL condition.
The state under test is extracted by the projection measurement onto V polarization from the output light of the 4-$f$ system.
The intensity measurements in time and frequency are realized by the time gate measurement and using an optical spectrum analyzer (Advantest Q8384), respectively.
The obtained temporal and spectral intensity distributions are used to examine the validity of the direct measurement results.

%%
% In this experimental setup, the time widths of the prepared test states were 
We note that the spectral width $\delta\omega$ extracted by the polarization-dependent frequency filter (1.08\,THz FWHM) is not sufficiently small compared to those of the states under test generated by the slit or the stripe mask ($\sim 6\,\mathrm{THz}$).
% The reference wave generated by the polarization-dependent frequency filtering was represented by the sinc function with the time width of about 6\,ps FWHM, which is not sufficiently large compared to that of the test states.
In this condition, the spectral wavefunction after the frequency filter should be approximated by the rectangle function $\mathrm{rect(\omega/\delta\omega)}$, which is zero outside the interval $[-\delta\omega/2,\delta\omega/2]$ and unity inside it.
In this case, the right sides of Eqs.~(\ref{eq:1}) and (\ref{eq:2}) are replaced by $\mathrm{sinc}(\delta\omega t/2)\mathrm{Re}[\psi\sub{env}(t)]$ and $\mathrm{sinc}(\delta\omega t/2)\mathrm{Im}[\psi\sub{env}(t)]$, respectively, where $\mathrm{sinc}(x):=\sin(x)/x$.
To obtain $\mathrm{Re}[\psi\sub{env}(t)]$ and $\mathrm{Im}[\psi\sub{env}(t)]$, we make a correction by dividing the measured wavefunctions by $\mathrm{sinc}(\delta\omega t/2)$, which is independent of $\psi\sub{env}(t)$ and was determined by prior measurement.
On the other hand, the time width of the gate pulses (79.2\,fs FWHM) is considered to be sufficiently smaller than those of the states under test ($\sim 3\,\mathrm{ps}$). 
Hence, we assume here that the effect of the width of the time measurement can be ignored.
The detailed calculation accounting for both the effects of the finite frequency and the time widths is given in Appendix~\ref{sec:appendixA}.
% 本セットアップにおいて, 被測定状態の時間幅は大体hoge\,psである. 
% 一方, 本実験系におけるスリットによる時間広がりはFWHMがXXXのsinc関数であり, 被測定状態の時間幅に対して無視できないので, その分を前述の通り補正した. 
% 一方, ゲートパルスの幅はhoge\,psなので, 被測定波動関数よりも十分小さいとして, 時間射影測定は正確に行われているとして考えた. 

% These widths of the time measurement and the frequency filter determine the temporal and spectral resolution of this direct measurement, respectively.

% 実際の測定系では, 偏光依存周波数フィルタ幅, 時間測定分解幅は有限の大きさを持つ. 
% これらの影響を考慮した計算は付録に記した. 
% 以下で述べるように, 我々の実験設定においては, 偏光依存周波数フィルタ幅の影響は無視できないが, 時間測定分解幅の影響は無視できると見積もった. 
% その場合, 周波数フィルタ後の周波数波動関数を$[\omega_0-\delta\omega/2,\omega_0+\delta\omega/2]$の矩形波と近似すると, Eqs.~(\ref{eq:1}), (\ref{eq:2})の右辺はそれぞれ$\mathrm{sinc}(\delta\omega t/2)\mathrm{Re}[\psi\sub{env}(t)]$, $\mathrm{sinc}(\delta\omega t/2)\mathrm{Im}[\psi\sub{env}(t)]$に置き換えられる. 
% したがって本実験では, 計測結果を事前測定した$\mathrm{sinc}(\delta\omega t/2)$の分布(これは被測定波動関数$\psi\sub{env}(t)$とは無関係な分布である)で割ることで$\mathrm{Re}[\psi\sub{env}(t)]$, $\mathrm{Im}[\psi\sub{env}(t)]$を得ている. 

%%%%%%%%%%%%%%%

\begin{figure}
\includegraphics[width=8.5cm]{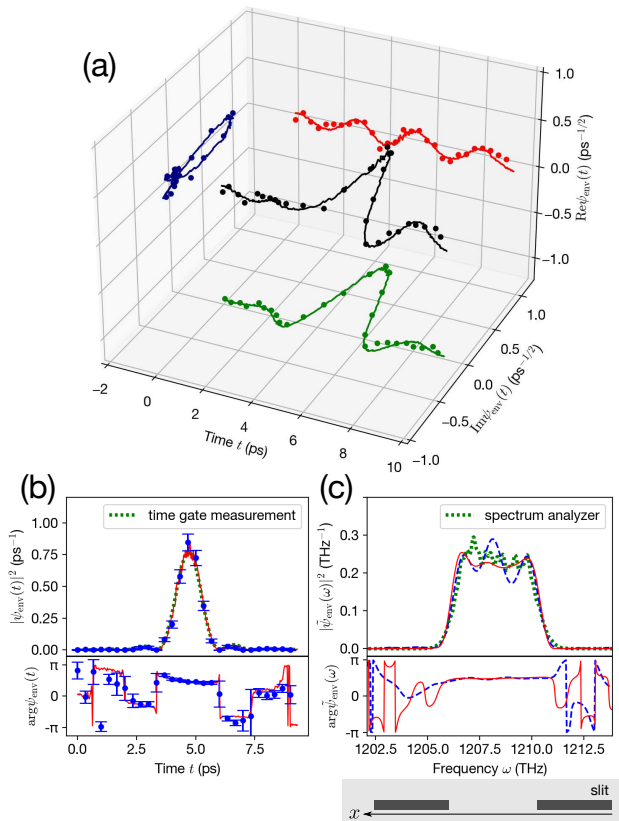}
\caption{
Results of the direct measurement of the wavefunction generated by the variable slit ($w=2.0$\,mm, $s=0.0$\,mm).
(a) 3D plot of the measured temporal wavefunction (black line and dots).
The red, green, and navy lines and dots are its projections on the real, imaginary, and amplitude phase planes, respectively.
The solid lines and the dots are the measurement results in the CL and SPL conditions, respectively.
In the SPL condition, photon counting was performed for 25 s per measurement point.
The error bars are omitted here.
(b) Intensity (upper panel) and phase distribution (lower panel) calculated from the measured temporal wavefunction.
The red solid line and blue dots are the measurement results in the CL and SPL conditions, respectively.
The error bars were calculated from the square root of the counted photon number (shot noise).
The green dotted line in the upper panel is the temporal intensity distribution obtained by the time gate measurement of the wavefunction under test.
(c) Intensity (upper panel) and phase distribution (lower panel) of the spectral wavefunction obtained by Fourier-transforming the measured temporal wavefunction.
The red solid line and blue dashed line are the distributions in the CL and SPL conditions, respectively.
The green dotted line in the upper panel is the spectral intensity distribution obtained using the optical spectrum analyzer for the wavefunction under test.
}\label{fig:3}
\end{figure}

In the following, we show the experimental results for state preparations (i)--(iii) in Fig.~\ref{fig:2}(c) in order.
% when a variable slit, a slit with a coverglass, and a stripe mask with two coverglasses are used as state preparations in order.
First, the spectral wavefunction generated by the variable slit with gap width $w$ and displacement $s$ is given by a rectangle function $\mathrm{rect}[(\omega-\omega\sub{c})/\Delta\omega]$.
The spectral width $\Delta\omega$ and central frequency $\omega\sub{c}$ are expressed as $\Delta\omega=\alpha w$ and $\omega\sub{c}=\alpha s$, respectively, where the proportional constant $\alpha:=2.41\,\mathrm{THz/mm}$ is derived from the geometrical configuration of our 4-$f$ system. 
The temporal wavefunction obtained by Fourier-transforming $\mathrm{rect}[(\omega-\omega\sub{c})/\Delta\omega]$ is $\ee^{\ii\omega\sub{c} t}\mathrm{sinc}(\Delta\omega t/2)$, and the time width $\Delta t$ between the two central zeros of this sinc function and the phase gradient $\kappa$ are given by $\Delta t=4\pi/\Delta\omega=4\pi/(\alpha w)$ and $\kappa=\omega\sub{c}=\alpha s$, respectively.
Therefore, in this state preparation, the form of the temporal wavefunction can be controlled quantitatively by changing $w$ and $s$.

We display the 3D plot of the result of the direct measurement of the temporal wavefunction generated by the variable slit ($w=2.0$\,mm, $s=0.0$\,mm) in Fig.~\ref{fig:3}(a).
There is no significant difference between the measurement results under the CL condition (lines) and those under the SPL condition (dots), while some fluctuation due to the shot noise is observed in the results in the SPL condition.
The intensity (square of the amplitude) and phase distributions of the measured temporal wavefunction are shown in Fig.~\ref{fig:3}(b), and those in the frequency domain, obtained by Fourier-transforming the measured temporal wavefunction, are shown in Fig.~\ref{fig:3}(c).
Furthermore, the temporal and spectral intensity distributions obtained by the time gate measurement and optical spectrum analyzer are displayed as green dotted lines in Figs.~\ref{fig:3}(b) and (c), respectively. 
The agreement of these intensity measurement distributions with the intensity distribution reconstructed from the directly measured wavefunction supports the validity of our direct measurement results.
A quantitative comparison between them using classical fidelity is discussed at the end of this section.

\begin{figure}
\includegraphics[width=8.5cm]{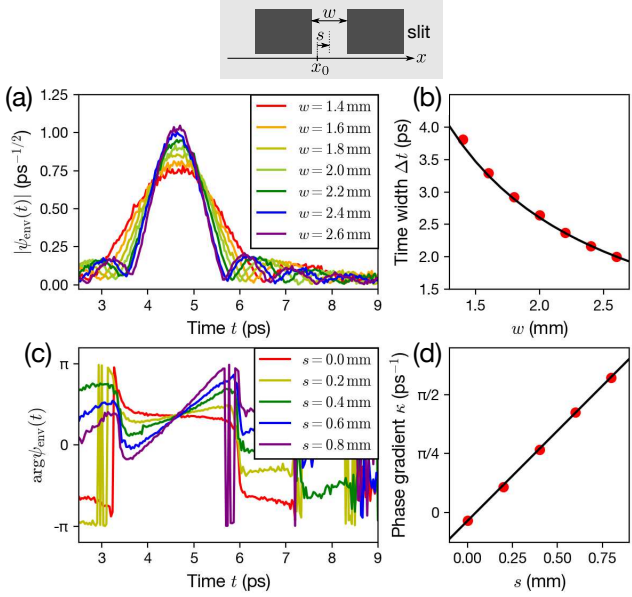}
\caption{Results of the direct measurement when the gap width $w$ and displacement of the gap center $s$ of the variable slit are changed.
(a) Measurement results of the magnitude of the temporal wavefunction when $w$ is changed from 1.4\,mm to 2.6\,mm while $s$ is fixed at $s=0\,\mathrm{mm}$.
(b) Relationship between $w$ and time width $\Delta t$ obtained from the measured curves.
The solid black line represents the theoretical curve $\Delta t=4\pi/(\alpha w)$.
(c) Measurement results of the phase of the temporal wavefunction when $s$ is changed from 0.0\,mm to 0.8\,mm while $w$ is fixed at $w=2.0\,\mathrm{mm}$.
(d) Relationship between $s$ and phase gradient $\kappa$ obtained from the measured curves.
The solid black line represents the theoretical curve $\kappa=\alpha s+\kappa_0$.
}\label{fig:4}
\end{figure}

Next, we examine the change in the measured temporal wavefunction when the gap width $w$ and displacement $s$ of the variable slit are changed.
% Next, we examined the validity of the direct measurement by observing the change in the measured temporal wavefunction when the gap width $w$ and the displacement $s$ of the variable slit are changed.
All these measurements are performed in the CL condition.
Figure~\ref{fig:4}(a) shows the direct measurement results of the magnitude of the temporal wavefunction when $w$ is changed from 1.4\,mm to 2.6\,mm while $s$ is fixed at $s=0\,\mathrm{mm}$. 
The time widths $\Delta t$ of the measured temporal amplitude, which are obtained by fitting the sinc function $A|\mathrm{sinc}[2\pi(t-t\sub{c})/\Delta t]|$ to the measured curves, are plotted versus $w$ in Fig.~\ref{fig:4}(b).
The values are in good agreement with the theoretical curve $\Delta t=4\pi/(\alpha w)$ (black line).
Figure~\ref{fig:4}(c) shows the direct measurement results of the phase of the temporal wavefunction when $s$ is changed from 0.0\,mm to 0.8\,mm while $w$ is fixed as $w=2.0\,\mathrm{mm}$. 
The phase gradients $\kappa$ of the measured temporal phase, which are also obtained by fitting the linear function to the measured curves in the range of $t\in[3.75\,\mathrm{ps},5.75\,\mathrm{ps}]$, are plotted versus the displacement $s$ in Fig.~\ref{fig:4}(d).
These values are also in good agreement with the theoretical curve $\kappa=\alpha s+\kappa_0$ (black line), where the offset value $\kappa_0:=-0.11\,\mathrm{ps}^{-1}$ is determined from the phase gradient when $s=0\,\mathrm{mm}$.

\begin{figure}
\includegraphics[width=8.5cm]{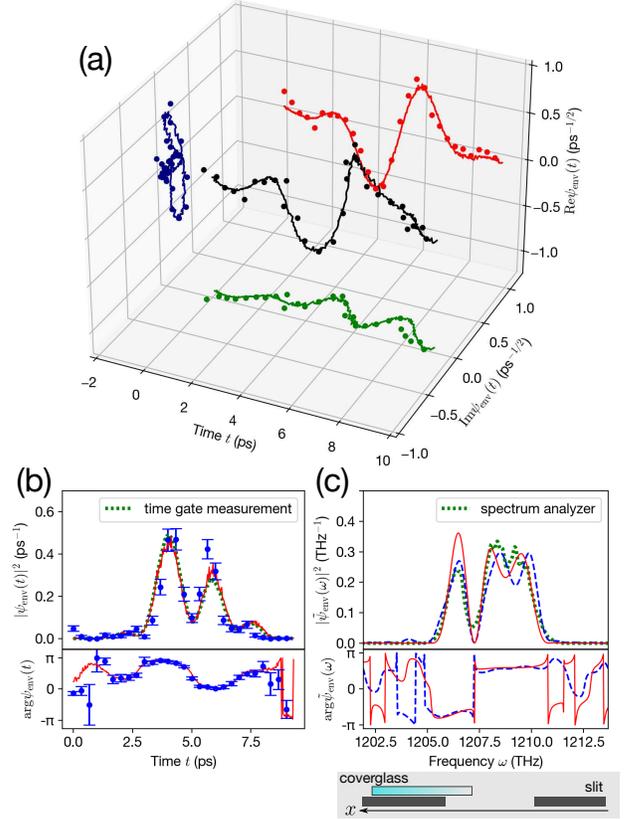}
\caption{
Results of the direct measurement of the temporal wavefunction generated by the slit ($w=2.0$\,mm, $s=0.0$\,mm) and coverglass.
The notation of this figure is the same as in Fig.~\ref{fig:3}.
% 状態生成にスリット($w=2.0$\,mm, $s=0.0$\,mm)とカバーガラスを用いた場合の波動関数の測定結果. 
% グラフの表記法は図~\ref{fig:3}と同様. 
}\label{fig:5}
\end{figure}

\begin{figure}
\includegraphics[width=8.5cm]{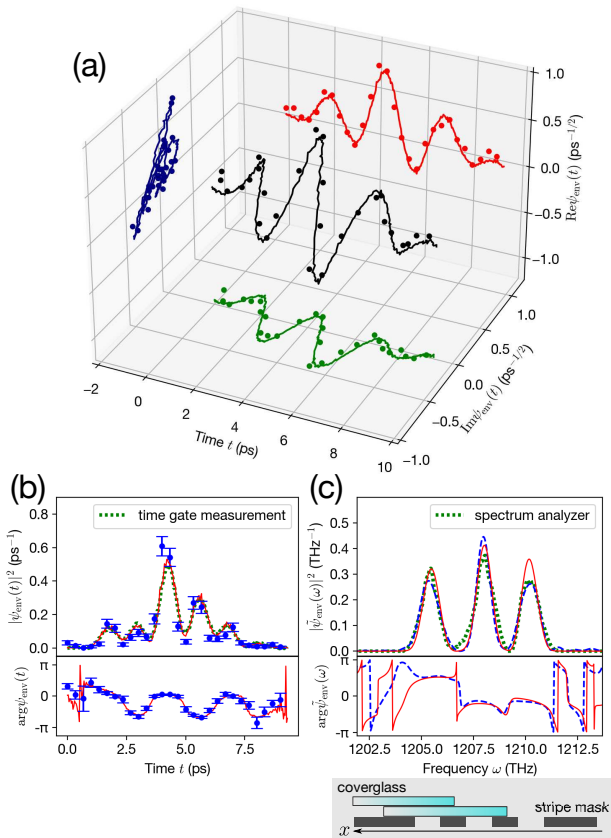}
\caption{
Results of the direct measurement of the temporal wavefunction generated by the stripe mask and coverglasses.
The notation of this figure is the same as in Figs.~\ref{fig:3} and \ref{fig:5}.
% 状態生成にストライプマスクとカバーガラスを用いた場合の波動関数の測定結果. 
% グラフの表記法は図~\ref{fig:3}と同様. 
}\label{fig:6}
\end{figure}

We further demonstrate the direct measurement of the temporal wavefunction generated by the slit ($w=2.0$\,mm, $s=0.0$\,mm) with a coverglass and by the stripe mask with two coverglasses.
The measurement results for the slit with a coverglass are shown in Figs.~\ref{fig:5}(a)--(c).
It should be noted that the frequency wavefunction derived from the directly measured time wavefunction shows a stepwise phase change due to the phase added by the coverglass.
The magnitude of the obtained phase step cannot be evaluated because its true value is not known in advance, as mentioned above.
Nevertheless, the agreement of the spectral intensity distributions derived from the directly measured time wavefunction (red and blue lines) with the results of the frequency intensity measurement (green line) indicates that the characterization of the wavefunction by direct measurement is performed properly. 
Figures~\ref{fig:6}(a)--(c) show the measurement results for the stripe mask with two coverglasses, which have more complicated waveforms. 
In this case as well, the point to be noted is that the frequency wavefunction derived from the directly measured time wavefunction (red and blue lines) shows two stepwise phase changes as a result of the two coverglasses, and their intensity distributions are in agreement with the results of the frequency intensity measurement (green line).
These results support the validity of the direct measurement method of the wavefunction.
 
\begin{table}
\caption{\label{tab:table1}
Classical fidelity (Bhattacharyya coefficient) between the intensity distributions calculated from the results of the direct measurement and those obtained by the projection measurements for panels (b) and (c) in Figs.~\ref{fig:3}, \ref{fig:5}, and \ref{fig:6}. 
CL and SPL indicate the signal power condition under which the direct measurements were performed.
}
\begin{ruledtabular}
\begin{tabular}{ccccc}
&\multicolumn{2}{c}{Time domain}&\multicolumn{2}{c}{Frequency domain}\\
&\multicolumn{2}{c}{[Panel (b)]}&\multicolumn{2}{c}{[Panel (c)]}\\
& CL & SPL & CL & SPL\\
\hline
Fig.~\ref{fig:3} & 0.999 & 0.993 & 0.995 & 0.990\\
Fig.~\ref{fig:5} & 0.998 & 0.973 & 0.985 & 0.974\\
Fig.~\ref{fig:6} & 0.999 & 0.976 & 0.987 & 0.970\\
\end{tabular}
\end{ruledtabular}
\end{table}

Finally, we evaluate the closeness of the intensity distributions of the wavefunctions obtained by the direct measurement and those obtained by the intensity (projection) measurement using the classical fidelity (Bhattacharyya coefficient).
The classical fidelity is defined as $\sum_j\sqrt{p_jq_j}$ for two probability distributions $\{p_j\}$ and $\{q_j\}$.
Table~\ref{tab:table1} shows the classical fidelity between the intensity distributions obtained by the direct measurement and the projection measurements for panels (b) and (c) in Figs.~\ref{fig:3}, \ref{fig:5}, and \ref{fig:6}. 
We can see that these fidelities show high values close to 1.

 % while $F\sub{SPL}$s are slightly less than one due to fluctuations in measured values due to the shot noise.

\section{Discussion}

First, we describe the performance of the direct measurement system used in our experiment.
The time resolution is determined by the time width of the gate pulse and the phase-matching bandwidth of SFG.
In our case, the latter effect is negligible and the time resolution is 79.2\,fs FWHM, which gives the subpicosecond resolution.
On the other hand, the measurable range in the time domain is determined by the time width of the self-generated reference light in the shape of a sinc function.
The time width between the two central zeros of the sinc function is 11.7\,ps.
Therefore, the dynamic range of our direct measurement system is evaluated to be $11.7\,\mathrm{ps}/79.2\,\mathrm{fs}=148$.

% ここでは、まず我々が構築した直接測定系の性能について述べる。
% 我々が構築した直接測定系では、時間分解能はゲートパルス光の時間幅とSFGの位相整合帯域のよって決まる。
% 我々の場合は後者の影響は無視できて、時間分解能は79.2\,fsであり、サブps分解能を達成した。
% 一方、時間領域における計測可能範囲は自己生成参照光のsinc関数の幅で決まり、中央の2つの零点の幅は11.7\,psであった。
% したがってこの測定系のダイナミックレンジは$11.7\,\mathrm{ps}/79.2\,\mathrm{fs}=148$と見積もられる。

% SFG + 空間周波数フィルタ + ファイバカップリング + ディテクタ量子効率 のこれらを含めた検出効率は$3.45\times10^{−4}$であった。

Next, we remark on previous studies related to direct measurement of the temporal wavefunction.
A recently reported experiment on $\updelta$-quench measurement \cite{PhysRevLett.123.190402} has demonstrated measurement of the temporal mode of light by applying instantaneous phase modulation followed by projection onto a specific frequency.
Although this method differs from direct measurement using weak measurement \cite{lundeen2011direct} and our direct measurement method, it satisfies the definition of direct measurement of the temporal wavefunction.
In this measurement, the time resolution did not reach the subpicosecond scale, and classical light much stronger than a single-photon level was used as the light under test.

In addition, a temporal-mode measurement method reported over 30 years ago \cite{rothenberg1987measurement} also satisfies the definition of direct measurement.
Although it was devised independently of the context of direct measurement, its configuration is similar to that of our direct measurement system. 
In this measurement, the time resolution reached the subpicosecond scale, while classical light was used as the light under test.
As a characterization method of the temporal mode of classical light, this method is currently rarely used in contrast to other sophisticated methods such as FROG and SPIDER.
%, which cannot be used for single photons.
However, the simple configuration of this method makes it suitable for the measurement of single photons, and the significance of our experiment is that it demonstrates this.

\section{Conclusion}

We proposed a direct measurement method for characterizing the temporal wavefunction of single photons and experimentally demonstrated the direct measurement for several test wavefunctions.
The experimental results showed that the direct measurement method works at the single-photon level and can achieve subpicosecond time resolution.
We clarified the validity of the direct measurement by quantitatively evaluating the measurement results when using the variable slit for state preparation and calculating the fidelities between the results of the direct measurement and the intensity distribution obtained by the projection measurement.
% 我々は光子の時間波動関数を特徴付ける手法として直接測定法を提案し, いくつかのテスト状態に対する測定実験のデモンストレーションを行った.
% 我々の測定では, 単一光子レベルでも測定可能であること, サブpsの時間分解能が達成できることを示した. 
% 我々は可変スリットを用いて準備されたテスト状態に対する測定結果の定量的評価, およびいくつかのテスト状態に対する直接測定の結果と射影測定により得られる強度分布とのフィデリティの評価を行い, この直接測定の妥当性を検証した. 

This direct measurement method can be applied not only to the temporal--spectral mode but also to other degrees of freedom.
In addition, it is expected that the direct measurement method can be extended not only to pure states but also to mixed states and processes; such an expansion of the scope of application of direct measurement is a subject for future research.

%it has been hoped to shed light on quantum gravity. 

\begin{acknowledgments}
This research was supported by JSPS KAKENHI Grant Number 19K14606, the Matsuo Foundation, and the Research Foundation for Opto-Science and Technology.
\end{acknowledgments}

%\clearpage

\appendix

\begin{widetext}
\section{Calculation of direct measurement method when resolution of frequency filter and time measurement is finite}\label{sec:appendixA}

Here, we describe the calculation of our direct measurement method when the effects of the finite resolution of the frequency filter and the time measurement are considered.
The projection operator of the frequency filter with spectral width $\delta\omega$ is given by $\int_{-\infty}^\infty\dd\omega\, \mathrm{rect}[(\omega-\omega_0)/\delta\omega]\ket{\omega}\bra{\omega}$, where $\mathrm{rect}[(\omega-\omega_0)/\delta\omega]$ is zero outside the interval $[\omega_0-\delta\omega/2,\omega_0+\delta\omega/2]$ and unity inside it.
The unnormalized resultant state after the polarization-dependent frequency filter is described as
\begin{align}
 \ket{\varPsi_1'}&=\frac{1}{\sqrt{2}}\left[\int_{-\infty}^\infty\dd\omega\, \mathrm{rect}\left(\frac{\omega-\omega_0}{\delta\omega}\right)\ket{\omega}\bracketi{\omega}{\psi}\ket{\HH}+\ket{\psi}\ket{\VV}\right].
\end{align}

The time measurement implemented by optical gating is characterized by the positive-operator-valued measure $\int_{-\infty}^\infty\dd t'g_t(t')\ket{t'}\bra{t'}$, where $g_t(t')$ is the non-negative gate function centered at $t'=t$.
The probability $P'(t,\phi)$ that the results of the time and polarization measurements are $t$ and $\phi$, respectively, is described as
\begin{align}
P'(t,\phi)=
\frac{\bra{\varPsi'_1}\left[\int_{-\infty}^\infty\dd t'g_t(t')\ket{t'}\bra{t'}\otimes\ket{\phi}\bra{\phi}\right]\ket{\varPsi'_1}}
{\bracketi{\varPsi_1'}{\varPsi_1'}}
=\int_{-\infty}^\infty\dd t'g_t(t')
\frac{\bracketii{\varPsi_1'}{(\ket{t'}\bra{t'}\otimes\ket{\phi}\bra{\phi})}{\varPsi_1'}}{\bracketi{\varPsi_1'}{\varPsi_1'}}.
\end{align}
Therefore, we obtain the following results:
\begin{align}
P(t,\DD)-P(t,\AA)
&=\int_{-\infty}^\infty\dd t'g_t(t')
\mathrm{Re}\left[\int_{-\infty}^\infty\dd\omega\, 
\mathrm{rect}\left(\frac{\omega-\omega_0}{\delta\omega}\right)
\bracketi{\phi}{\omega}\bracketi{\omega}{t'}\bracketi{t'}{\psi}\right],\\
P(t,\RR)-P(t,\LL)
&=\int_{-\infty}^\infty\dd t'g_t(t')
\mathrm{Im}\left[\int_{-\infty}^\infty\dd\omega\, 
\mathrm{rect}\left(\frac{\omega-\omega_0}{\delta\omega}\right)
\bracketi{\phi}{\omega}\bracketi{\omega}{t'}\bracketi{t'}{\psi}\right].
\end{align}

Assuming that $\bracketi{\omega}{\psi}$ is the constant value $\bracketi{\omega_0}{\psi}$ in the interval $[\omega_0-\delta\omega/2,\omega_0+\delta\omega/2]$, the integral with respect to $\omega$ can be calculated as  
\begin{align}
\int_{-\infty}^\infty\dd\omega\, 
\mathrm{rect}\left(\frac{\omega-\omega_0}{\delta\omega}\right)
\bracketi{\phi}{\omega}\bracketi{\omega}{t'}
=\frac{\bracketi{\psi}{\omega_0}}{\sqrt{2\pi}}
\ee^{\ii\omega_0 t'}\delta\omega
\mathrm{\,sinc}\left(\frac{\delta\omega t'}{2}\right),
\end{align}
and then we obtain
\begin{align}
P(t,\DD)-P(t,\AA)
&\propto
\int_{-\infty}^\infty\dd t'g_t(t')
\mathrm{\,sinc}\left(\frac{\delta\omega t'}{2}\right)
\mathrm{Re}[\psi\sub{env}(t')],\\
P(t,\RR)-P(t,\LL)
&\propto
\int_{-\infty}^\infty\dd t'g_t(t')
\mathrm{\,sinc}\left(\frac{\delta\omega t'}{2}\right)
\mathrm{Im}[\psi\sub{env}(t')].
\end{align}
Furthermore, when the temporal width of the optical gate is sufficiently small compared with that of $\psi\sub{env}(t)$, we can approximate $g_t(t')=\delta(t-t')$ and thus obtain
\begin{align}
P(t,\DD)-P(t,\AA)
&\propto
\mathrm{\,sinc}\left(\frac{\delta\omega t}{2}\right)
\mathrm{Re}[\psi\sub{env}(t)],\quad
P(t,\RR)-P(t,\LL)
\propto
\mathrm{\,sinc}\left(\frac{\delta\omega t}{2}\right)
\mathrm{Im}[\psi\sub{env}(t)].
\end{align}
We adopt these approximated results in the main text.
% In this case, $\mathrm{Re}[\psi\sub{env}(t)]$ and $\mathrm{Im}[\psi\sub{env}(t)]$ are obtained by dividing them by $\mathrm{sinc}(\delta\omega t/2)$, which is independent of $\psi\sub{env}(t)$ and can be determined by prior measurement.

% \section{Geometrical configuration of the 4-$f$ system}

% ここでは我々が実験で構成した4-$f$系の構成の詳細を述べ、Eq.~(\ref{eq:3})の関係式を導出する. 
% 図に我々が構成した4-$f$系を図示する。

% 図5では得られた波動関数からフィッティングにより求めた特徴量である時間幅$\Delta t$と位相勾配$a$を、系の幾何学的構成から決まる理論曲線と比較した。
% ここではそのフィッティングの具体的な方法と、理論曲線の導出法について述べる。

% フィッティング関数はこんな式。
% このbがわかれば、そこから時間幅Dtが求められる。
% 同じく、位相勾配も同様。

% 理論曲線の導出：
% 図のような構成になっている。
% この時、回折格子で回折される光の周波数と角度の関係を導出。
% 次に角度と位置の関係を導出。
% よって周波数と位置の関係を導出。

% 周波数矩形波の幅とsinc関数の時間幅Dtの関係を説明、
% ここから、Dtとwの関係がわかる。
% 同じく、シフト量と位相勾配の関係から、aとsの関係がわかる。

% We fitted the function $a|\mathrm{sinc}[b(t-c)]|$ to the measured curves, and obtained their time widths (FWHM) as $\Delta t=3.79/b$.
% Theoretically, the relation between the time width $\Delta t$ and the gap width $w$ is derived as $\Delta t=3.16\,\mathrm{ps}(\mathrm{mm}/w)$ from the geometrical configuration of the 4-$f$ system.

% We fitted the function $at+b$ to the measured curves in the range of $t\in[3.75\,\mathrm{ps},5.75\,\mathrm{ps}]$ and obtained the phase gradients $a$.
% Theoretically, the relation between the phase gradient $a$ and the displacement $s$ is derived as $a=(2.40s/\mathrm{mm}-0.11)\,\mathrm{ps}^{-1}$ from the geometrical configuration of the 4-$f$ system, where the offset value $-0.11$ was determined from the phase gradient when $s=0\,\mathrm{mm}$.

\end{widetext}

\end{spacing}

\nocite{*}
\bibliography{ref}% Produces the bibliography via BibTeX.

\end{document}